\documentclass[prd,showpacs,amsmath,showkeys,floatfix,amssymb, preprintnumbers, nofootinbib, superscriptaddress]{revtex4} 
\usepackage{hyperref} 
\usepackage{amsmath,amssymb,bm}
\usepackage{epsfig}
\usepackage{graphicx,color}
\usepackage{amsfonts}
\usepackage{epstopdf}
\usepackage{cases}
\usepackage[capitalise]{cleveref}

\newcommand\lb{\left(}
\newcommand\rb{\right)}
\newcommand\ls{\left[}
\newcommand\rs{\right]}
\newcommand\lc{\left\{}
\newcommand\rc{\right\}}

\renewcommand{\part}{{\rm part}}

\newcommand{\ket}[1]{\left| #1\right\rangle }

\newcommand{\matelem}[3]{\left\langle #1\left| #2 \right| #3\right\rangle}

\newcommand{\nn}{\nonumber}
\newcommand{\beqa}{\begin{eqnarray}}
\newcommand{\eeqa}{\end{eqnarray}}

\begin{document}

\title{Emergent Kink Statistics at Finite Temperature}

\author{Miguel Angel Lopez-Ruiz}
\email{malopezr@indiana.edu}
\affiliation{ Physics Department and Center for Exploration of Energy and Matter,
Indiana University, 2401 N Milo B. Sampson Lane, Bloomington, IN 47408, USA.}

\author{Tochtli Yepez-Martinez}
\email{yepez@fisica.unlp.edu.ar}
\affiliation{ Physics Department and Center for Exploration of Energy and Matter,
Indiana University, 2401 N Milo B. Sampson Lane, Bloomington, IN 47408, USA.}  
\affiliation{Departamento de Fisica, Universidad Nacional de La Plata, C.C.67 (1900), La Plata, Argentina.}

\author{Adam Szczepaniak}
\email{aszczepa@indiana.edu}
\affiliation{ Physics Department and Center for Exploration of Energy and Matter,
Indiana University, 2401 N Milo B. Sampson Lane, Bloomington, IN 47408, USA.}
\affiliation{Thomas Jefferson National Accelerator Facility, Newport News, VA 23606, USA.}

\author{Jinfeng Liao}
\email{liaoji@indiana.edu}
\affiliation{ Physics Department and Center for Exploration of Energy and Matter,
Indiana University, 2401 N Milo B. Sampson Lane, Bloomington, IN 47408, USA.}
\affiliation{RIKEN BNL Research Center, Bldg. 510A, Brookhaven National Laboratory, Upton, NY 11973, USA.}

\date{\today}

\begin{abstract}
In this paper we use 1D quantum mechanical systems with Higgs-like interaction potential to study the emergence of topological objects at finite temperature. Two different model systems are studied, the standard  double-well potential model and a newly introduced discrete kink model. Using Monte-Carlo simulations as well as analytic methods, we demonstrate how kinks become abundant at  low temperatures. These results may shed useful insights on how topological  phenomena may occur in QCD. 
\end{abstract}
\maketitle

\section{Introduction} 
\label{Sec: introduction}

Emergent topological objects are known to exist and play important role in various physical systems. Well know examples include e.g. instantons and magnetic monopoles in Yang-Mills theories or vortices in superfluids and superconductors~\cite{Shuryak-book,Coleman,'tHooft:1999au,Schaf-Shuryak,Bruckmann,Tong:2005un,Gani:2015cda,Gani:2014gxa,Gani:1998jb}. In Quantum Chromodynamics (QCD) it is widely expected that the key nonperturbative properties, confinement and chiral symmetry breaking are driven by topological objects, with magnetic monopoles or vortices responsible for the former and instantons for the latter~\cite{'tHooft:1999au,Schaf-Shuryak,Bruckmann}. Numerous studies of the QCD vacuum and hadron properties have been performed along these lines (see e.g.~\cite{Shuryak:1981ff,Shuryak:1984nq,Shuryak:1993kg,Greensite:2011zz,Szczepaniak:2010fe,Szczepaniak:2003ve,Szczepaniak:2001rg}). There is also significant interest in understanding QCD at finite temperature. When heated, the normal hadronic matter turns into a hot phase of QCD, known as the quark-gluon plasma (QGP)~\cite{Shuryak-book}. Experimentally the quark-gluon plasma is created in high energy heavy ion collisions  at the Relativistic Heavy Ion Collider and the Large Hadron Collider. As shown by lattice QCD simulations, such a transition occurs at temperatures near $T_c \sim 165\rm MeV$, above which hadrons melt. Remarkably the so-obtained QGP, with a temperature range around $1\sim 4 T_c$, is found to be strongly coupled~\cite{Shuryak:2004cy,Gyulassy:2004zy}. Many of its key transport properties are found to bear highly nonperturbative  that are likely to be  related to emergent topological objects already present and dominant for QGP in the near-$T_c$  regime~\cite{Liao:2006ry,Liao:2008jg,Liao:2008dk,Liao:2012tw,Shuryak:2008eq,D'Alessandro:2007su,Bonati:2013bga,D'Alessandro:2010xg}. 
   
Many interesting studies have been performed to describe the QCD transitions and the QGP properties at finite temperature, by postulating existence of a statistical ensemble of various topological objects. A well-known example is the Instanton Liquid Model (ILM, see e.g.~\cite{Schaf-Shuryak}) which lead to a rather accurate description of the spontaneous breakdown of chiral symmetry in the vacuum as well as the chiral restoration at finite temperature. Recently these efforts have been extended to include ensembles of calorons (i.e. finite temperature instantons) that have nontrivial holonomy (see e.g.~\cite{Lopez-Ruiz:2016bjl,Larsen:2015vaa,Larsen:2015tso,Liu:2015ufa,Liu:2015jsa}) and such models seem to be able to simultaneously describe both the confinement and chiral properties of QCD systems. To understand how ensembles of topological objects may emerge from the first principle QCD, remains a significant challenge. If indeed nonperturbative dynamics can be captured by an ensemble of such objects, then it is of fundamental importance to derive their dynamics from the full QCD partition function and investigate their statistical properties. In this paper we aim to address this question in a simplified model. By understanding the emergence of statistical ensemble of topological objects in a model one may hope to gain useful insights on how the same question might be addressed for QCD. To this end, we will use 1D quantum mechanical systems with Higgs-like interaction potential. For such a system  a type of tunneling solution known as the kink is the topological object and it is a close analog of the instanton in QCD.  Two different models will be discussed: the well known standard double-well potential model will be treated through Monte Carlo Metropolis (MCM) simulations as comparison tool for a new discrete-kink model (DKM) here developed, which allows approximate analytic solutions of the partition function with explicit kink statistics. To the best of our knowledge, this DKM has not been studied in the past. We will systematically study such statistical systems at finite temperature using numerical simulations and analyze to what extent key properties of such systems may be understood via emergent kink statistics. 

The rest of this paper is organized as follows. In Sec. II we review the standard double-well potential model to introduce the topological objects of interest, namely the kink (and anti-kink), to then introduce the new discrete kink model. Details of numerical studies are given in Sec. III together with a comparison of the analytic predictions of the new discrete kink model and an analysis in terms of kink statistics.

\section{The Discrete Kink Model approach for the Double-Well Potential}

In this section, before delving into the Discrete Kink Model (DKM), we first give a brief discussion of the physics of kinks, the topological objects in 1D quantum mechanical systems with a double-well potential, characterized by being localized finite energy minima of the Euclidean action at strong coupling. For many years, the standard treatment of such systems has relied on semiclassical methods; here, we present a novel discrete approach at finite temperature which allows to obtain exact analytic expressions for the partition function and therefore, different observables such as Green's functions.

\subsection{Kink in the standard double-well potential model}\label{DWKink}

\begin{figure}[]
\centering
\includegraphics[width=0.7\textwidth]{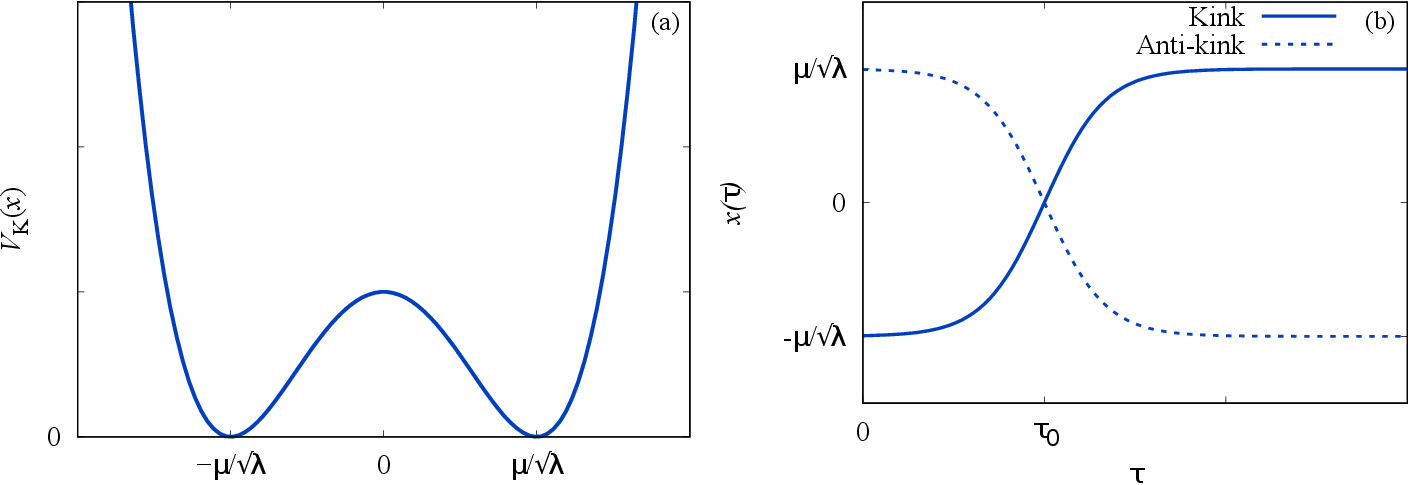}
\caption{(a) Standard double-well potential.  (b) Kink and anti-kink solutions.}
\label{Fig:kink-pot&sol}
\end{figure}

We begin with a 1D quantum mechanical system described by the standard double-well potential given by 

\begin{equation} \label{Eq:kink-potential}
V_{\text{K}}(x) = \frac{\lambda_k}{4} \left(x^2-\frac{\mu_k^2}{\lambda_k}\right)^2,
\end{equation} 
which has two stable minima at $x_{min}= \pm \mu_k/\sqrt{\lambda_k}$ (Fig.~\ref{Fig:kink-pot&sol}(a)). Here $\lambda_k$ is the coupling constant with mass dimension $[\lambda_k]\sim [m]^5$ and a VEV parameter $\mu_k$ with mass dimension $[\mu_k] \sim [m]^{3/2}$. Suppose the particle mass parameter (in the kinetic energy term) is $m$, we will then work with dimensionless quantities rescaled by proper powers of $m$ (e.g.  $\mu_k^2/m^3\rightarrow\mu_k^2$ and $\lambda_k/m^5\rightarrow\lambda_k$) throughout this paper. The potential (\ref{Eq:kink-potential}) is depicted in Fig.~\ref{Fig:kink-pot&sol}(a). 

We focus on properties of the system at finite temperature $T$, which can be studied through the partition function,

\begin{equation}
\label{Eq:partition-function}
\mathcal{Z}\equiv\mbox{Tr}\left(e^{-\beta H}\right)=\int\limits_{x(\beta) = x(0)}\mathcal{D}x\,e^{-S_E[x(\tau)]}, 
\end{equation} 
where $k_B=\hslash=1$, $\beta=1/T$ is the inverse of temperature $T$, $H$ is the Hamiltonian and $S_E$ is the Euclidean action given by 

\begin{equation}
S_E[x(\tau)]= \int_0^\beta d\tau {\cal L}_E(x(\tau)) = 
\int_0^\beta \mbox{d}\tau\left[\dfrac{1}{2}\left(\dfrac{\mbox{d}x}{\mbox{d}\tau}\right)^2 + V_{\text{K}}(x)\right]   \,\, .
\end{equation}

The Lagrangian ${\cal L}_E$ corresponds to that of a classical particle moving along the $x$ axis subject to the potential $-V_{\text{K}}(x)$ \cite{Bruckmann,Coleman}. The finite energy minima of $S_E$ correspond to classical trajectories of such analog particle. This requires  $-V_{\text{K}}(x)$ to vanish as $\tau \rightarrow \pm \infty$,  i.e $x(\tau\rightarrow\pm\infty)=\pm\mu_k/\sqrt{\lambda_k}$.  In this case, the equations of motion have the two trivial solutions $x(\tau)=\pm\mu_k/\sqrt{\lambda_k}$ corresponding to the analog particle in an unstable equilibrium in one of the maxima of $-V_{\text{K}}(x)$. There are also two nontrivial solutions given by 

\begin{equation}
x(\tau)=\pm\frac{\mu_k}{\sqrt{\lambda_k}}\tanh\left[\frac{\mu_k}{\sqrt{2}}(\tau-\tau_0)\right],
\label{Eq:kink-solution}
\end{equation}
describing a rolling motion of the analog particle form one maximum of $-V_{\text{K}}(x)$ to the next over infinite long time. The above solutions with plus and minus sign are respectively called the kink  and anti-kink  solutions.  The two solutions are depicted in Fig.~\ref{Fig:kink-pot&sol}(b) and the parameter $\tau_0$ marks its location or position along the $\tau-$axis. If one translates the solution   back to the original Minkowski formulation, the kink solutions can be interpreted as the quantum tunneling of the physical, quantum  particle through the potential barrier from one of the minima of $V(x)$ to the other.  It shall be noted that the kink in double-well potential model is of course a standard example with   many excellent expositions in the literature (see e.g. \cite{Bruckmann,Coleman,Schaf-Shuryak,Shuryak-book,Polyakov:1976fu,abcInstantons,Rajaraman,Razavy}), and what we include here is just a brief discussion for reader's convenience. 

Kinks are localized objects with size $\Delta\tau\sim 1/\omega$, where $\omega=\ls V''(\pm\mu_k/\sqrt{\lambda_k})\rs^{1/2}=\sqrt{2}\mu_k$ is the frequency of small oscillations around each minimum. In the region $\beta = 1/T\gg \Delta\tau$, configurations consisting of chains of kinks and anti-kinks are favorable to appear and will contribute the most to the partition function $\mathcal{Z}$, as we shall see in the numerical analysis presented in \cref{Sec:MCMvsDKM}. The increasing number of kinks contributing at low temperatures suggests that physical observables can be analyzed by considering the emergent statistical system of an ensemble of such kinks, which is the main motivation in the model presented in the following.

\subsection{A Discrete-Kink Model}
\label{Sec:DKM}

The discrete-kink model (DKM) is defined on a discrete temporal lattice in Euclidean time, with $\tau_i$   $(i=-N+1,...,0,...,N+1)$ being the discretized Euclidean points spanning the interval $[0, \beta]$. $2N+1$ is the total number of lattice points, and $a=\beta/(2N+1)$ is the lattice spacing. The position variable $x_i=x(\tau_i)$ on each lattice site is a discrete version of the $x(\tau)$ in the continuum. We introduce a  potential $V_{\text{DKM}}(x_i)$ by, 

\begin{equation} 
V_{\text{DKM}}(x_i) = -\frac{1}{a}   \log   \lc   \exp   \ls    
-a \mu(a)^2   \lb   x_i-\frac{\mu(a)}{\sqrt{\lambda(a)}}   \rb^2   \rs   
+\exp   \ls    
-a \mu(a)^2   \lb   x_i+\frac{\mu(a)}{\sqrt{\lambda(a)}}   \rb^2   \rs  \rc
+C(a)~.
\label{Eq:V-DKM}
\end{equation} 

We note that the above potential explicitly depends on the parameter $a$ with the parameters $\mu$ and $\lambda$ also depending on $a$. For any given value of $a$, these parameters can be suitably adjusted to make  $V_{\text{DKM}}(x_i)$ as close as possible to the original double-well  potential  $V_{\text{K}}(x)$ from  Eq.~(\ref{Eq:kink-potential}).  The explicit $a$-dependence of the parameters of the 
potential $V_{\text{DKM}}(x_i)$ will be given later.  The new potential is similar to the double-well potential, namely  there are two minima located at $x\approx\pm \mu/\sqrt{\lambda}$, which support the kink/anti-kink solutions. Thus this model mimics the standard double-well potential model in a discrete version limit and the advantage of this seemingly complicated form is that it will allow an analytic treatment of the partition function, as we show below.

The partition function of Eq.~(\ref{Eq:partition-function}) can be expressed in a discretized form as
\begin{equation}
\mathcal{Z}=\int\prod_{j=-N+1}^{N+1}\left(\dfrac{\mbox{d}x_j}{\sqrt{2\pi a}}\right)\exp\left\lbrace -\sum_{i=-N+1}^{N+1} \left[\dfrac{\left(x_{i+1}-x_i\right)^2}{2a} +aV(x_i)\right]\right\rbrace.
\label{Eq:Path-Int-Disc}
\end{equation}

Note that  periodic boundary conditions are imposed such that $x_{N+1}\equiv x_{-N}, x_{N+2}\equiv x_{-N+1}$, etc. We have also rescaled quantities with proper dimension of mass, i.e. $am\rightarrow a$ and $mx\rightarrow x$. Introducing the source term, $\mathcal{Z} \to \mathcal{Z}[j]$ and replacing 
 $V(x_i)$ by the potential $V_{\text{DKM}}(x_i)$, we obtain the generating functional of the DKM given by
 
\beqa
\mathcal{Z}[j] 
&=& 
\int      \frac{ \mbox{d}x_{N+1} }{ \sqrt{2\pi a} }   
   \frac{ \mbox{d}x_{N} }{  \sqrt{2\pi a} }
\cdots       \frac{ \mbox{d}x_{-N+1} }{ \sqrt{2\pi a} } 
\prod_{i=-N+1}^{N+1}  
\exp   \ls   -    
\lb   \frac{ (x_i -x_{i-1})^2 }{ 2a}  + a j_i x_i +aC(a)
\rb   
\rs~ 
\nn\\
&\times&  \lc \exp   
\ls   -  a\mu (a)^2  \lb   x_i - \frac{\mu (a) }{ \sqrt{ \lambda (a) } }   \rb^2      \rs + \exp   
\ls   -  a\mu (a)^2  \lb   x_i + \frac{\mu (a) }{ \sqrt{ \lambda (a) } }   \rb^2      \rs
\rc 
\nn \\
&=& 
\int   \sum_{n_{N+1}=\pm1}   \frac{ \mbox{d}x_{N+1} }{ \sqrt{2\pi a} }   
\sum_{n_{N}=\pm1}   \frac{ \mbox{d}x_{N} }{  \sqrt{2\pi a} }
\cdots   \sum_{n_{-N+1}=\pm1}   \frac{ \mbox{d}x_{-N+1} }{ \sqrt{2\pi a} }
\nn\\
&\times&   \exp   
\lc   -\sum_{i=-N+1}^{N+1}   
\ls   \frac{ (x_i -x_{i-1})^2 }{ 2a}
+a\mu (a)^2    \lb   x_i - \frac{n_i \mu (a) }{ \sqrt{ \lambda (a) } }   \rb^2   
+a j_i x_i +aC(a)
\rs   
\rc ,
\label{Eq:DKM_action}
\eeqa
with $2N+1$ integrations and $2N$ intermediate states. In the above, one could recognize the advantage brought in by the logarithmic form in the definition of the potential Eq. (\ref{Eq:V-DKM}). In particular
 one can explicitly recognize the two terms  at each lattice site, with one corresponding to picking up the potential around one minimum and the other corresponding to picking up the potential around the other minimum. We next develop a method to explicitly compute this generating functional.

\subsubsection{Integration via change of variables}

The action in Eq. (\ref{Eq:DKM_action}) has  a quadratic form  and can be diagonalized by
a change of variables $x_n$, $n=-N\cdots,N \to \mbox{Re}\,b_q, \mbox{Im}\,b_q$,
$q=1,\cdots,N$, $b_0 (\mbox{Im}\,b_0 = 0)$ i.e. going over to the
conjugate lattice with $q's$ defined on links, 

\beqa\label{Eq:change-variables}
x_n  &=&  \sum_{q=-N}^{N}  \frac{e^{2\pi i  \frac{nq}{2N+1} }}{\sqrt{2N+1}} b_q 
=   \frac{1}{\sqrt{2N+1} }  \ls   b_0   
+\sum_{q=1}^N   e^{ 2\pi i  \frac{nq}{2N+1} }b_q
+\sum_{q=1}^N   e^{ -2\pi i  \frac{nq}{2N+1} }b^*_q \rs \nn\\
&=&   \frac{1}{\sqrt{2N+1} }  \ls   b_0   
+2\sum_{q=1}^N   \cos \lb  \frac{ 2\pi nq}{2N+1}   \rb   \mbox{Re}\,b_q
-2\sum_{q=1}^N   \sin \lb  \frac{ 2\pi nq}{2N+1}   \rb   \mbox{Im}\,b_q \rs,
\eeqa
with $b_{-q} = b^*_q$. 

The Jacobian is computed from

\beqa
&&\frac{\partial (x_N,\cdots x_N)}{\partial (b_0, \mbox{Re}\,b_1,\cdots \mbox{Re}\,b_N,
  \mbox{Im}\,b_1, \cdots \mbox{Im}\,b_N )}= 2^N,
\eeqa
so that

\beqa
\int \mbox{d}x_{-N} \cdots \mbox{d}x_N= \int \mbox{d}b_0 \left[2\mbox{d}(\mbox{Re}\,b_1)\, \mbox{d}(\mbox{Im}\,b_1)\right]\cdots \left[2\mbox{d}(\mbox{Re}\,b_N)\,
\mbox{d}(\mbox{Im}\,b_N)\right]
=\int   \mbox{d}b_{-N} \cdots \mbox{d}b_N ~.
\eeqa

Then, defining 

\beqa
\tilde{j}_q=\sum\limits_{n=-N}^N\frac{e^{-2\pi i\frac{nq}{2N+1}}}{\sqrt{2N+1}}j_n, \quad \tilde{n}_q=\sum\limits_{m=-N}^N\frac{e^{-2\pi i\frac{mq}{2N+1}}}{\sqrt{2N+1}}n_m \quad\text{and}\quad \tilde{q}=2\sin \lb \frac{\pi q}{2N+1}\rb,
\eeqa 
with $\tilde{j}_{-q}=\tilde{j}_q^*$  and $\tilde{n}_{-q}=\tilde{n}_q^*$, the action becomes 

\beqa
&&\sum_{i=-N+1}^{N+1} 
\ls  \frac{(x_i -x_{i-1})^2}{2a} +a\mu^2 \lb  x_i -\frac{n_i
  \mu }{\sqrt{\lambda }}  \rb^2 +aj_ix_i      \rs   \nn\\
&&=(2N+1)a\frac{\mu^4}{\lambda} 
+ a\sum_{q=-N}^N \lb   \frac{\tilde{q}}{2a^2} +\mu^2   \rb
\lb b_q^* +\frac{  -\frac{\mu^3}{\sqrt{\lambda} } \tilde{n}_q^* + \frac{\tilde{j}_q^*}{2} } 
  {  \frac{\tilde{q}^2}{2a^2} +\mu^2 } \rb
\lb b_q +\frac{  -\frac{\mu^3}{\sqrt{\lambda} } \tilde{n}_q + \frac{\tilde{j}_q}{2} } 
  {  \frac{\tilde{q}^2}{2a^2} +\mu^2 } \rb
-a\sum_{q=-N}^N  
\frac{  \left| -\frac{\mu^3}{\sqrt{\lambda} } \tilde{n}_q^* +  \frac{\tilde{j}_q^*}{2} \right| } 
  {  \frac{\tilde{q}^2}{2a^2} +\mu^2 },  \nn\\
\eeqa
where  the parameters $\mu$ and $\lambda$ of the potential are still $a$-dependent. After performing the integrations the generating functional becomes:

\beqa \label{Eq_Z_j}
\mathcal{Z}[j] 
&=& \sum_{n_{N+1}=\pm 1}  \cdots \sum_{n_{N+1}=\pm 1}
\exp \ls
-\frac{1}{2}  \sum_{q=-N}^N \log \left[\tilde{q}^2 +2a^2 \mu^2\right]
+ a\sum_{q=-N} ^N   
\lb  -\frac{\mu^4}{\lambda}  + 
\frac{  \left|-\frac{\mu^3}{\sqrt{\lambda} } \tilde{n}_q^* +  \frac{\tilde{j}_q^*}{2} \right| } 
  {  \frac{\tilde{q}^2}{2a^2} +\mu^2 }     \rb   \rs \nn.\\
\eeqa

Let us first handle the  term $\log \lb\tilde{q}^2 +2a^2 \mu^2\rb$. Note that in the continuum limit from the soluble harmonic oscillator (identifying  harmonic term $m^2 \to 2a^2\mu^2$ and $\tilde{q}/a \to k =2\pi q/[(2N+1)a] =2 \pi q /\beta$), one has the following result, 

\beqa
-\frac{1}{2} \sum_{q=-N}^N \frac{1}{\tilde{q}^2 +m^2}
&\approx&-\frac{1}{2} \sum_{q=-N}^N \frac{1}{\frac{(2\pi)^2 q^2}{(2N+1)^2}+m^2}
=-\frac{(2N+1)^2}{2(2\pi)^2} 
\sum_{q=-N}^N \frac{1}{q^2 +\frac{(2N+1)^2m^2}{(2\pi)^3} }\nn\\
&=&-\frac{(2N+1)^2}{2(2\pi)^2}  \frac{2\pi^2}{(2N+1)m} 
\coth \ls \frac{(2N+1)m}{2} \rs
=-\frac{2N+1}{4 m}   \coth \ls \frac{(2N+1)m}{2} \rs.
\eeqa

The above can be related to the  term $\log \lb\tilde{q}^2 +2a^2 \mu^2\rb$ by the following relation 

\beqa
-\frac{1}{2} \frac{\mbox{d}}{\mbox{d}m^2} \sum_{q=-N}^N \log \lb\tilde{q}^2 +m^2\rb_{m^2=2\mu^2a^2}
=-\frac{2N+1}{4 m}   \coth \ls \frac{(2N+1)m}{2} \rs.
\eeqa 

Thus by integrating the above over variable $m$, we obtain: 

\beqa\label{A:harmonic-term}
&&-\frac{1}{2} \sum_{q=-N}^N \log\lb\tilde{q}^2 +m^2\rb 
= -\frac{2N+1}{2}   \int \mbox{d}m \coth\ls  \frac{(2N+1)m}{2} \rs +\mbox{const.} \nn\\
&& =-\frac{2N+1}{2}  \ls  -m -\frac{2 \log (2) }{ 2N+1 }
-\frac{2}{2N+1} \log \lb \frac{1}{e^{(2N+1)m}-1} \rb  \rs
= \frac{(2N+1)m}{2} +\log(2) +\log \lb
\frac{e^{-(2N+1)m}}{1-e^{-(2N+1)m}}\rb +\mbox{const.}\nn\\
&&=\frac{\beta\mu}{\sqrt{2}} +\log \lb
\frac{e^{-\beta \sqrt{2} \mu  }}{1-e^{-\beta \sqrt{2} \mu } }\rb
+\log(2) +\mbox{const.}
= \log \lb \frac{e^{-\beta \sqrt{2} \mu\frac{1}{2}  }}{1-e^{-\beta
    \sqrt{2} \mu } }\rb
=\log\lb \sum_{n=0}^\infty  e^{-\beta\sqrt{2} \mu (n+\frac{1}{2})}\rb.
\eeqa

Note the additional constant contributes the same multiplicative coefficient to all terms in the generating functional Eq. (\ref{Eq_Z_j}) and can be safely dropped. We next  compute the  kinetic  term in the action, 

\beqa\label{A:propagator-term}
a\sum_{q=-N}^N \frac{\tilde{n}_q^*  \tilde{n}_q}{\frac{\tilde{q}^2}{2a^2}  + \mu^2}
&=&a^2 \sum_{n,m=-N}^N n_n  \ls \sum_{q=-N}^N \frac{1}{(2N+1)a}  
\frac{e^{-2\pi i \frac{(an-am)q}{(2N+1)a}}}{\frac{\tilde{q}^2}{2a^2}
  +\mu^2}  \rs  n_m \nn \\
&=& a^2 \sum_{n,m=-N}^N n_n G(n - m) n_m,
\eeqa
where we introduce a site propagator $G(n - m)$, given as follows in the limit $\sqrt{2}\mu\beta\gg 1$ and $N\gg 1$: 
 
\beqa
 G(n - m)= \frac{2\beta}{(2\pi)^2}  
\sum_{q=-\infty}^\infty 
\frac{e^{-2\pi i q \frac{(n-m)a}{\beta} } }{q^2 
  +\frac{ 2\mu^2 \beta^2} {(2\pi)^2} }
\approx\frac{1}{\sqrt{2}\mu}   e^{-|n-m| a\sqrt{2}  \mu }  \quad .
\eeqa 

Finally, substituting the above results into the generating functional Eq. \eqref{Eq_Z_j}, we get the following results: 

\beqa\label{Eq:Z-before-kink}
\mathcal{Z}[j] &=& \lb  \sum_{n=0}^\infty e^{-\beta \sqrt{2} \mu  (n+\frac{1}{2})} \rb
\sum_{n_{N+1}=\pm 1} \cdots \sum_{n_{-N+1}=\pm 1} 
e^{-\beta \frac{\mu^4}{\lambda}}\nn\\
&\times&\exp   
\lc a^2 \sum_{n,m=-N}^N    \ls   
\lb  \frac{\mu^3}{\sqrt{\lambda}} \rb^2    n_n G(n -m) n_m
-\lb \frac{\mu^3}{\sqrt{\lambda}} \rb  n_n G(n -m) j_m  
+   \frac{1}{4}  j_n G(n -m) j_m  \rs  \rc.\nn\\
\eeqa

\subsubsection{Emergent kink statistics}

\begin{figure}[hbt!]
\begin{center}
\includegraphics[width=0.4\textwidth]{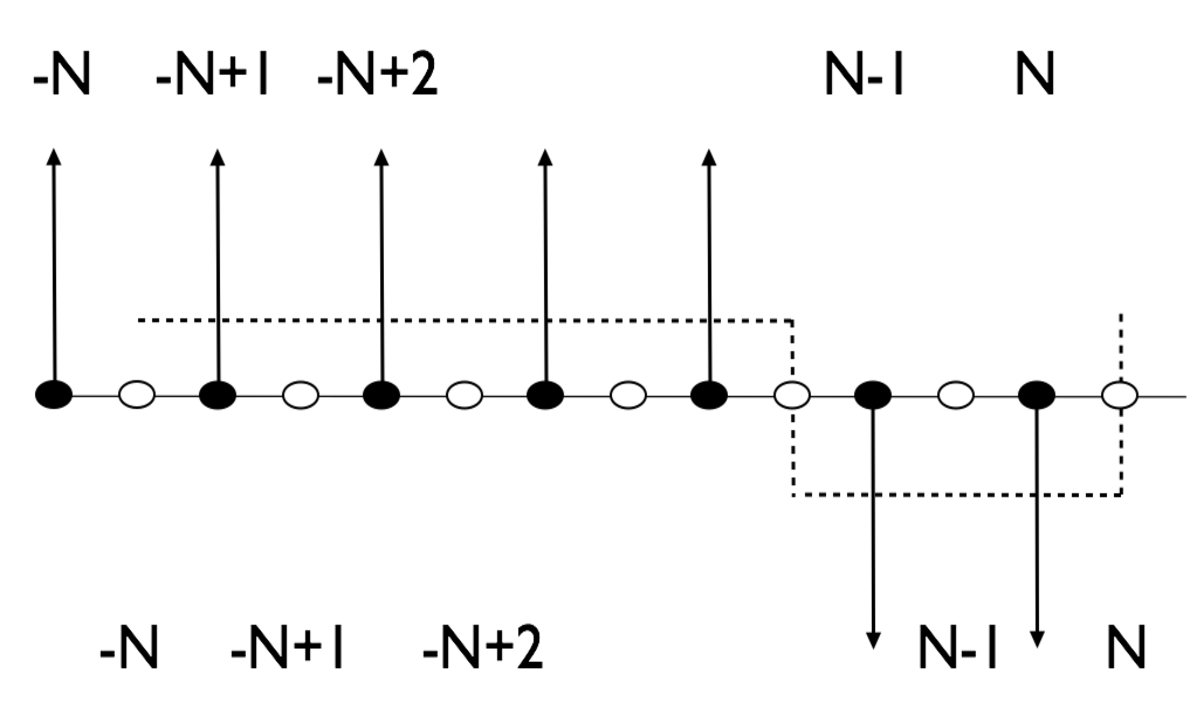}
\end{center}
\caption{Example of $n_i$ distribution on the N=3 lattice. Vertical
  arrows indicate the $n_i = \pm 1$ variables located on the lattice
  sites $i=-N,\cdots N$ represented by solid circles. Kinks are
  denoted by dashed lines and are represented on lattice links,
  represented by open  circles. Link (site) at $ N+1$ is identified with the link (site) at $-N$.} \label{Fig:fig-lattice}
\end{figure}

To complete the evaluation of the partition function, the last step is to do the sum  over $n_i 's$ in  Eq.~(\ref{Eq:Z-before-kink}). Recall that the variable $n_i=\pm 1$, which may be called a site-index, chooses the location of the particle to be one or the other minimum of the  potential. Consider two neighboring sites $x_i$ and $x_{i+1}$: if $n_i=n_{i+1}$ the configuration corresponds to the situation of the particle staying at the same potential minimum i.e. absence of a kink/anti-kink transition between the two sites. If, however $n_i=- n_{i+1}$ the configuration corresponds to the situation of the particle staying at two opposite potential minima i.e. with a kink or anti-kink tunneling occurring in this time step. In fact, each term in the sum, with a specific set of $n_i$'s, can be equivalently specified by specifying whether there is a kink or anti-kink, on each link between two neighboring sites. This is illustrated in  Fig.~\ref{Fig:fig-lattice}.  For example, a configuration with all site-indexes $n_i$ equally plus or minus one, corresponds to a configuration with no kinks and no anti-kinks at any link. A configuration with all site-index equal except  for one, e.g, $n_j=1$ with all  other $n_{i\neq j}=-1$, corresponds to a kink configuration with one kink and one anti-kink at link $(j-1)\to j$ and link $j \to (j+1)$ respectively. It is easy to see how such correspondence generalizes to more complicated situations.  Given this correspondence one can  replace the sum over sites by a sum over links with all possible kink/anti-kink/none configurations. We therefore see the explicit emergence of kink statistics in this model. 

Having converted   the summation over $n_i$'s into a sum over kink configurations we can organize all possible kink configurations according to the number of kinks in each configuration. Note that due to periodic boundary conditions, the number of kinks must equal the number of anti-kinks in the same configuration. So we classify all possibilities into  configurations with 0-kink 0-anti-kink, 1-kink 1-anti-kink, 2-kink 2-anti-kink$\cdots$, etc. As the technical derivation of this kink-anti-kink summation is quite lengthy and involved, we defer it to the \cref{Sec:DKM-Kink-Counting}. Here we simply present the final analytic result for the partition function $\mathcal{Z}=\mathcal{Z}[j_i \to 0]$, 

\beqa
\mathcal{Z} = \lb   \sum_{n=0}^\infty e^{-\beta\sqrt{2} \mu
  (n+\frac{1}{2})  }  \rb
\ls  1 + \frac{\beta^2}{2!a^2}F^2  + \frac{\beta^4}{4!a^4}F^4 +
\frac{\beta^6}{6!a^6}F^6 +\cdots   \rs
=\lb   \sum_{n=0}^\infty e^{-\beta \sqrt{2} \mu (n+\frac{1}{2})  }   \rb
\cosh \lb \sqrt{2} \mu \beta F\rb,   
\label{Eq:DKM-Z}
\eeqa
with $ F = \exp\lb-\sqrt{2} \mu^3/\lambda\rb  $. The corresponding  correlation function $\langle x(\tau)x(0) \rangle$ is also obtained as follows: 

\beqa
\langle x(\tau) x(0) \rangle =\frac{G(\tau)}{2} +\frac{\mu^2}{\lambda} 
\frac{\cosh \lb (\beta-2\tau)\sqrt{2} \mu F\rb}{\cosh \lb \sqrt{2} \mu \beta F\rb}  \,\, .
\label{Eq:DKM-CF}
\eeqa

From the partition function Eq. \eqref{Eq:DKM-Z}, the average number of kinks/anti-kinks can also be calculated analytically for the DKM

\beqa
\langle N_k \rangle&=&\frac{1}{\mathcal{Z}} \sum_{n=0}^\infty  e^{-\sqrt{2}\mu\beta(n+\frac{1}{2}) } 
\sum_{N_k=0,2,\cdots}^{\infty} \frac{N_k}{N_k!} (\sqrt{2}\mu \beta
F)^{N_k}
=\sqrt{2} \mu \beta ~\tanh
( \sqrt{2} \mu \beta ).
\label{Eq:DKM-Nk}
\eeqa

From the above one can easily read the statistical distributions of configurations with given number of kinks/anti-kinks

\beqa
f(N_k ) = \frac{1}{\cosh(\xi)} \frac{\xi^{N_k} }{N_k!}, 
\label{Eq:DKM-Dist}
\eeqa
with $\xi=\sqrt{2} \mu \beta F$. At low temperature one has approximately $f(N_k) \approx 2\xi^{N_k} e^{-\xi} / N_k!$ which is the well known Poisson distribution, with the additional factor of two accounting for presence of both kinks and anti-kinks. It is not surprising that the latter distribution is consistent with that of an ideal gas; since the approximation $\sqrt{2}\mu\beta\gg 1$, essential for the derivation of the partition function, suggests that at sufficiently small temperatures the Euclidean time scale is much larger than the kink size $\Delta\tau$, thus enabling configurations with several well defined kinks and anti-kinks which do not overlap with each other, resembling a non-interacting gas of kinks.

Let us conclude this  section by emphasizing the difference between the discrete-kink Model (DKM)  here and the well known instanton gas model~\cite{Polyakov:1976fu}.  Usually in such semi-classical approach as the instanton gas, the starting point is the individual instanton/anti-instanton (or kink/anti-kink) which is the individual classical solution, and then one assumes an ensemble of such objects as an approximation to the exact partition function. In the DKM study here, the particular functional form of the ``double-well''-like potential allows to compute the partition function analytically from summing contributions of  multi-``kink-like" configurations which naturally emerge from the above derivation.

\section{Numerical results of Standard kink potential vs DKM}
\label{Sec:MCMvsDKM}

\begin{figure}[]
\centering
\begin{minipage}[b]{0.45\textwidth}
\includegraphics[width=\textwidth]{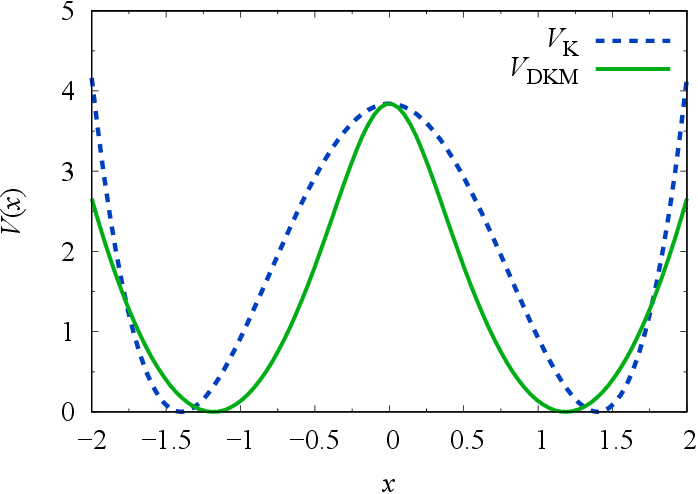}
\caption{Standard double well potential and the DKM potential.}
\label{Fig:Pot}
\end{minipage}
\hspace{0.5cm}
\begin{minipage}[b]{0.45\textwidth}
\includegraphics[width=\textwidth]{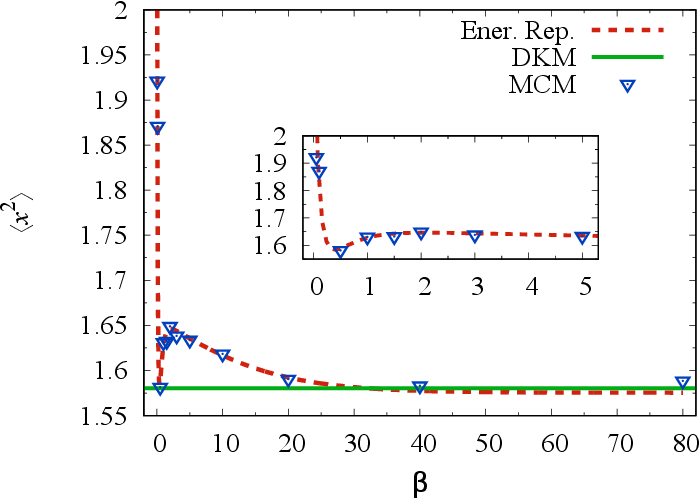}
\caption{Expectation value $\left\langle x^2 \right\rangle$ from MCM of standard kink potential and the DKM analytic result.}
\label{Fig:X2}
\end{minipage}
\end{figure}

In this section we discuss the numerical simulation of the standard kink potential, how their results can be interpreted through the emergent kink statistics and compare them to those of the DKM. To begin with, let us give some technical details of the numerical calculations. We first note that not all parameters in the potential \cref{Eq:kink-potential} are independently influencing the correlation function and it is possible to simplify the simulations by fixing, $\lambda_k$ and $\mu_k$ as suggested in \cite{Schaefer}. In particular, we use $\lambda_k=4$ and $\mu_k/\sqrt{\lambda_k}=1.4$. In addition, since the main purpose of the numerical analysis of the standard kink potential, is to test the analytic results obtained in the previous section with the DKM, we choose a suitable set of parameters $\left\lbrace \mu,\lambda,a,C \right\rbrace$ which define a potential $V_{\text{DKM}}$ as similar as possible to $V_{\text{K}}$ and are given by $\mu=2,\, \lambda=2.85,\, a=0.391$ and $C=4.105\times10^{-4}$ (see \cref{Fig:Pot}). For the Monte Carlo simulation, smaller lattice sizes require larger number of steps before convergence is achieved and the simulations for different temperatures $(T=1/\beta)$ were done with a different number of sweeps ranging from  $N_{MC}=10^8$ for $\beta=80$ to $N_{MC}=6\times10^9$ for $\beta=0.05$.  

Let us first discuss the results for $\left\langle x^2\right\rangle$ at low temperatures. As shown in \cref{Fig:X2}, in the low-$T$ regime, $\left\langle x^2\right\rangle$ is almost flat and gradually grows with increasing temperature. From the perspective of energy representation, this is simply due to the fact that the leading contribution comes from the ground state that dominates the low-$T$ behavior of the system. Regarding the DKM result, from Eq. \eqref{Eq:DKM-CF} we see that $\left\langle x^2\right\rangle=1/2\sqrt{2}\mu + \mu^2/\lambda\approx 1.5803$, which is temperature independent and its numerical value lies between those of the energy representation (see \ref{Eq:EnerRep-CF}) and the MCM, with satisfactory accuracy. 

\begin{figure}[]
\centering
\includegraphics[width=0.6\textwidth]{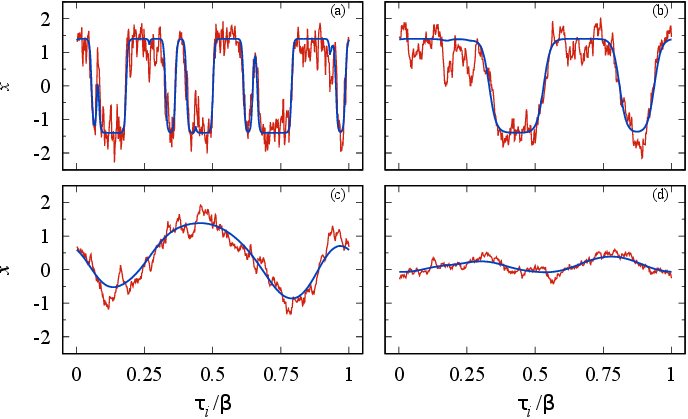}
\caption{Monte Carlo configurations $\left\lbrace x_i\right\rbrace$ for (a) $\beta=80$, (b) $\beta=20$, (c) $\beta=5$ and (d) $\beta=1$. The red and blue curves correspond to the ``raw" and ``cooled" configurations respectively. }
\label{Fig:PHI4-Trajectories}
\end{figure}

To understand the temperature independence of the DKM result one has to look into the interpretation in the language of the path integral formulation (see \cref{Eq:PathRep-CF}) which is more complicated and also interesting as it demonstrates the role of the kinks. To see that, let us examine explicitly sample configurations of the quantum particle locations, $\left\lbrace x_i \right\rbrace$ extracted from the MCM simulation at various temperatures, as shown in \cref{Fig:PHI4-Trajectories}. At the lowest temperature (\cref{Fig:PHI4-Trajectories}(a)), one can easily recognize several well-defined kink-like and anti-kink-like trajectories (where the particle suddenly ``jumps'' from position near one of the potential minima to the other even within a single configuration. As temperature increases one observes that there appears fewer sharp kink-like trajectories and that the individual kink-like trajectories  become ``smoother'' and less ``recognizable''.  

We plot a one sample configuration at each temperature, but these observations are quite general. These features can be qualitatively understood as follows. As mentioned previously, kinks and anti-kinks have a narrow but finite size $\Delta\tau$ (i.e. the time span of the quantum tunneling process). An increase in temperature  reduces  the span of the total imaginary time interval $[0,\beta]$, leaving less time available to accommodate kinks and anti-kinks and thus suppressing their number. Furthermore, the kinks and anti-kinks are classical solutions to the equation of motion and subject to thermal fluctuations. Such fluctuations will smooth out the sharp kinks as they become more important as temperature increases. 

Given that the degrees of freedom in the DKM are kink-like objects, we look for the kink content in the system with the standard kink potential $V_{\text{K}}$ in order to show how in the low temperature regime, these objects are in fact the main contributors to the thermodynamics of the system, thus justifying the kink statistics to describe it.

To approach this in a quantitative way, we have performed a statistical analysis with a sample of $10^4$ MCM configurations at each temperature. The number of kinks/anti-kinks is counted in each configuration to generate statistical distributions of the kink content. An accurate analysis, requires an accurate counting method for the number of kinks/anti-kinks per configuration, denoted by $N_k$. As it can be seen in \cref{Fig:PHI4-Trajectories}, crossings through the horizontal $x$-axis could be a way of determining $N_k$. However, a naive counting over the raw MCM configurations will give a number that is larger than the actual number of kinks/anti-kinks due to thermal fluctuations.
 
In order to get reliable statistics, a cooling method ~\cite{Schaefer,Hoek1,Hoek2} was used. The method consists in applying once again the MCM algorithm (see \cref{Sec:monte-carlo-method}) to an already accepted configuration $\left\lbrace x_i \right\rbrace$ accepting configurations which minimize the action. This reduces quantum fluctuations and helps to isolate the kink content allowing for a more efficient counting of kinks per configuration. For instance, in \cref{Fig:PHI4-Trajectories} we see that the raw MCM sample (in red) can have small fluctuations when $x(\tau_i)\approx 0$, which may cause over-counting  but is avoided after cooling procedure (in blue). The cooling, however, has to be used carefully since configurations that have sharp peaks, i.e. a kink and an anti-kink with a small separation, can be smoothed out to a point where it no longer counts as kinks crossing. To avoid this, one has to tune the number of cooling sweeps so that it is large enough to get rid of fluctuations around the crossings, but small enough to be able to count important crossings.

\begin{figure}[]
\centering
\includegraphics[width=1\textwidth]{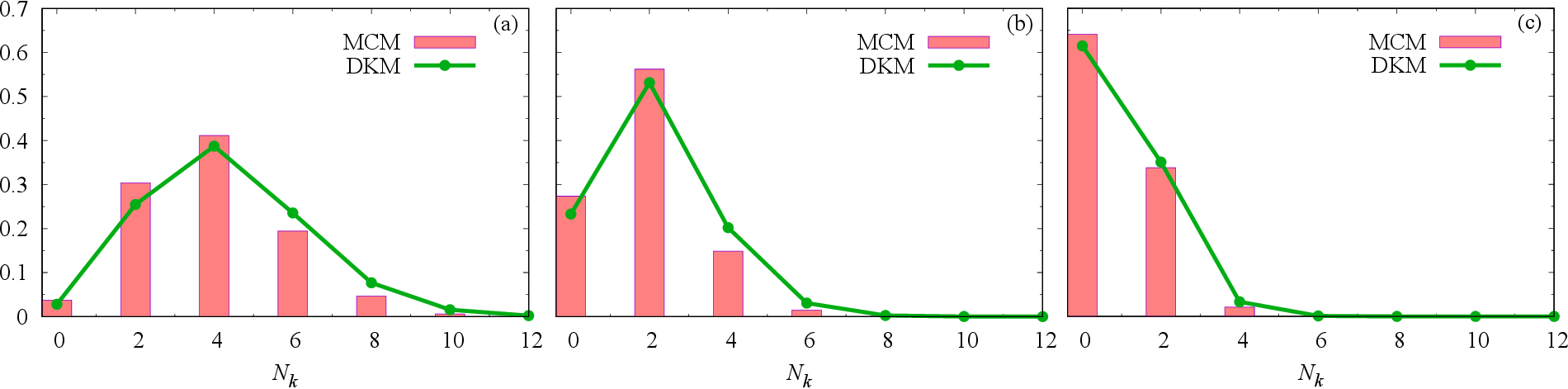}
\caption{Normalized distribution of $N_k$ for (a) $\beta=80$, (b) $\beta=40$ and (c) $\beta=20$.}
\label{Fig:PHI4-Hist}
\end{figure} 

In \cref{Fig:PHI4-Hist} we show the distribution of $N_k$ for the lowest temperatures simulated ($\beta=80,40$ and 20). On top of the histograms, we show the predicted kink distribution of the DKM (Eq. \eqref{Eq:DKM-Dist}), which is seen to agree reasonably well with the simulation results. As discussed previously, we then show that indeed in the very low temperature regime, configurations containing finite number of kinks are more favorable while at higher temperature configurations without kinks/anti-kinks gradually become preferred. 

\begin{figure}[]
\centering
\includegraphics[width=1.0\textwidth]{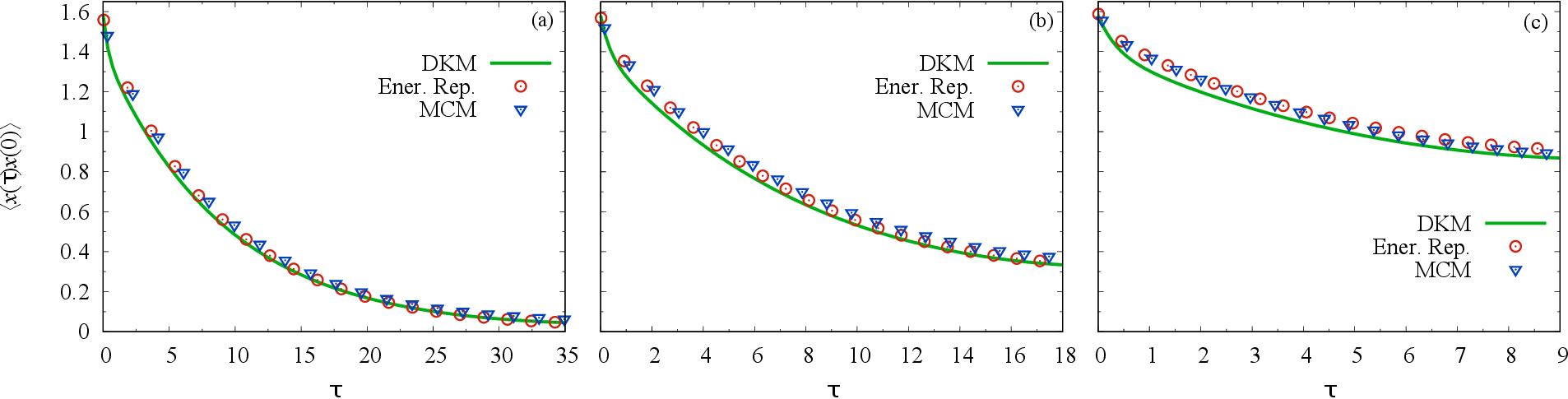}
\caption{Correlation functions computed in the energy representation and MCM compared to the DKM result for (a) $\beta=80$, (b) $\beta=40$ and (c) $\beta=20$.}
\label{Fig:PHI4-CF}
\end{figure}

We now look at the correlation functions $\left\langle x(\tau)x(0) \right\rangle$, shown in \cref{Fig:PHI4-CF}. The DKM result of Eq. \eqref{Eq:DKM-CF} agrees quite well with those of the MCM and energy representation. Qualitatively, we can say that a kink or anti-kink (``jumping'' from positive to negative minimum) will make a negative contribution to this correlation function. Indeed, for lower temperatures one sees a strongly suppressed tail of the correlation function as compared to the one at higher temperatures where the correlation function becomes flatter, indicating that the configurations become more dominated by fluctuations rather than by kinks.   

Lastly, motivated by the abundance of kinks in the low-$T$ regime, we examine whether the MCM results can be reproduced by a simple approximate model of a statistical ensemble based on kink/anti-kink degrees of freedom. In such approximation, each configuration $x(\tau)$ (for $\tau \in [0,\beta]$) is approximated by a sum of segments: those without kink/anti-kink events with the particle staying at some average position $\pm x_0$ and those where kink/anti-kink configurations. The kinks/anti-kinks have a finite time span of $\tau_k$ and the average of $x^2$ over this time span is denoted by $x_k^2$ which is a constant determined by the detailed shape of kink/anti-kink. In this simple picture, the observable $\left\langle x^2 \right\rangle$ for a configuration with $N_k$ of kinks/anti-kinks, is given by 

\begin{equation}\label{Eq.linear-model}
\left\langle x^2 \right\rangle \left(N_k\right) = \frac{\int_0^\beta x(\tau)^2 d\tau}{\beta} \approx 
\frac{x_0^2 (\beta - N_k \tau_k) + x_k^2 (N_k \tau_k)}{\beta} 
= x^2_0 - \frac{\left(x^2_0 - x_k^2\right)\tau_k}{\beta}N_k.
\end{equation}

\begin{figure}[]
\centering
\includegraphics[width=1.0\textwidth]{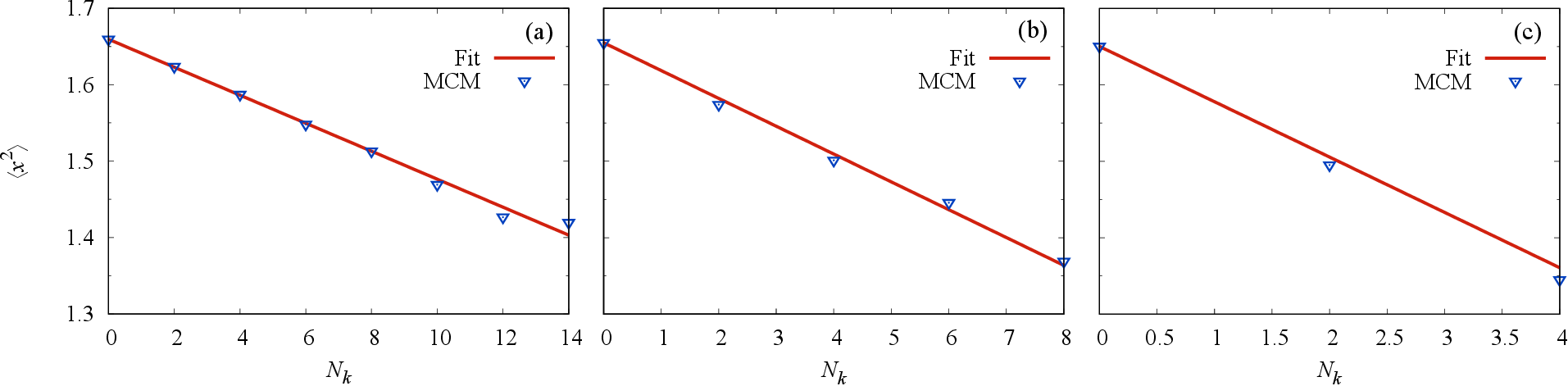}
\caption{Model of $\left\langle x^2\right\rangle$ for a given kink number $N_k$ for (a) $\beta=80$, (b) $\beta=40$ and (c) $\beta=20$.}
\label{Fig:PHI4-x2Ave-Model}
\end{figure}

The above approximation predicts a specific linear dependence of $\left\langle x^2 \right\rangle$ on $N_k$. It should be emphasized that the parameter $x_0^2$ is temperature dependent due to thermal fluctuations, while the kink parameters $\tau_k$ and $x_k^2$ both pertain only to kink properties and are temperature independent. To verify this approximation we compute $\left\langle x^2 \right\rangle$  from configurations with given kink/anti-kink numbers and  analyze its dependence on $N_k$ at  different temperatures ($\beta=80, 40$ and $20$), see \cref{Fig:PHI4-x2Ave-Model}. An excellent linear dependence is seen at these temperatures. The fit of the $N_k$ dependence at different $\beta$ confirms the temperature independence of the kink parameters, yielding  $x_k^2\approx 0.87$ and $\tau_k \approx 1.85$. Therefore we have demonstrated that in the low temperature regime where thermal fluctuations are not significant while quantum tunneling is important, a statistic ensemble of kinks/anti-kinks provide a very good description to the double-well potential model. While the double-well potential model was studied numerically in the past (see e.g. \cite{Schaefer,Alexander:1993ns,Habib}), the above interpretation of $\left\langle x^2 \right\rangle$ via our kink statistics analysis appears to be new. 
 
\section{Summary} 

In this paper we used 1D quantum mechanical system with Higgs-like interaction potential to study the emergence of topological objects at finite temperature. We have developed a discrete kink model, which allows for analytic computation at finite temperature. For this new model, we have explicitly derived the partition function in terms of summation over kink/anti-kink configurations, providing a clear picture of the emergence of kink statistics in this system. The analytic results are compared with numerical simulations of the standard kink potential similar to that of the discrete model and found to be in outstanding agreement with each other. From both, the analytic and numeric studies, we conclude that kinks/anti-kinks become abundant in the low temperature regime where the behavior of the system is well described by a statistical ensemble. Our findings have provided a thorough illustration of how topological objects like the kinks can emerge in a statistic system and dominate the system's properties. These results may shed useful insights on how similar phenomenon may occur in more complicated systems such as the Yang-Mills theories at finite temperature.

\vskip0.2cm

\section*{Acknowledgments}

The authors are grateful to S. Gottlieb for very helpful discussions on the numerical methods. The research of MALR and JL is supported by the National Science Foundation (Grant
No. PHY-1352368).  JL is also grateful to the RIKEN BNL Research
Center for partial support. MALR is in addition supported by CONACyT
under Doctoral supports Grants No. 669645. 
TYM thanks for Postdoctoral supports to CONACyT/Mexico under Grants No. 166115 and
No. 203672 and CONICET/Argentina. 
AS is supported in part by U.S. Department of Energy, Office of Science, Office of Nuclear Physics under contract DE-AC05-06OR23177 and by the U.S. Department of Energy under Grant No. DE-FG0287ER40365.  The computation of this research was performed on IU's Big Red II cluster that is supported in part by Lilly Endowment, Inc. (through its support for the Indiana University Pervasive Technology Institute) and in part by the Indiana METACyt Initiative.


\appendix 

\section{Monte Carlo Simulation} 
\label{Sec:monte-carlo-method}

For a statistical system like the one given in Eq. (\ref{Eq:partition-function}), analytic calculation of correlation functions is most often not possible and we have to rely on numerical tools.  The Metropolis algorithm \cite{Metropolis} for Monte Carlo numerical integration (MCM) has proven to be extremely useful for evaluation of the path integral. In the present study, the observable of interest is the two point correlation function  $\left\langle x(\tau)x(0)\right\rangle$, defined as 

\begin{equation} \label{Eq:PathRep-CF}
\left\langle x(\tau)x(0)\right\rangle=\mathcal{Z}^{-1}\mbox{Tr}\left[x(\tau)x(0)e^{-\beta H}\right]=\dfrac{\int\mathcal{D}x\,x(\tau)x(0)e^{-S_E[x(\tau)]}}{\int\mathcal{D}x\,e^{-S_E[x(\tau)]}},
\end{equation}
together with its value at the origin $\langle x^2 \rangle =\langle x(0)x(0) \rangle$. To numerically compute the correlation function we use an ensemble of configurations  $\left\lbrace x_i^{(k)}\right\rbrace$ of the system generated and distributed according to the Boltzmann distribution $e^{-S_E[x_i]}/\mathcal{Z}$. Note that the temporal direction $\tau\in [0,\beta]$ is discretized into an equally-spaced ``lattice'' with $i=1,\ldots,N$ labeling the lattice sites. The index $k=1,\ldots,N_{MC}$ labels the sequence of each configurations from Monte Carlo ``sweeps" in the simulation. This process begins with a random initial configuration $x_i^{(0)}$ and is updated to a new configuration as $x_i^{(1)}= x_i^{(0)}+\delta x$, where $\delta x$ is a random number. The Metropolis criteria states that the new configuration will be accepted if $S_E[x_i^{(1)}]<S_E[x_i^{(0)}]$, otherwise $x_i^{(1)}$ is accepted with probability $P(x_i^{(0)}\rightarrow x_i^{(1)})=e^{-(S_E[x_i^{(1)}] - S_E[x_i^{(0)}])}$. This is done for all $2N+1$ points of the lattice for a total number of $N_{MC}$ sweeps until the system has been stabilized, i.e. when the average value of the observable has converged to some number up to a fixed precision criteria.  To obtain a reliable simulation, we use a $50\%$ acceptance rate on configuration updates. To ensure this, we adjust the width of the distribution of our random displacement $\delta x$, which depends on the value of temporal lattice spacing parameter $a$ used in the simulations. For our calculations we have fixed the number of lattice points to $2N+1=500$, and computed the 2-point correlation function for different values of  $\beta$, ranging between 0.05 and 80. 

As a test of the MCM simulation, we computed $\left\langle x(\tau)x(0)\right\rangle$ using the energy representation and compared the results from the two methods. In the energy representation, the correlation function  is given by
 
\begin{equation}
\left\langle x(\tau)x(0)\right\rangle= \dfrac{\sum\limits_{n,m}e^{E_n(\tau-\beta)}e^{-\tau E_m}\left|\matelem{n}{x(0)}{m}\right|^2}{\sum\limits_{n}e^{-\beta E_n}}.
\label{Eq:EnerRep-CF}
\end{equation}

Here the eigenstates $\ket{n}$ and energy eigenvalues $E_n$ are not analytically. We compute them  by numerically solving the Schr\"odinger's equation with the accurate Numerov's method \cite{Numerov1}. For this calculation, a total of 30 lowest eigenstates and eigenvalues were used.

\section{Detailed derivation of the kink-anti-kink summation}
\label{Sec:DKM-Kink-Counting}

In this Appendix we present the detailed derivation of the kink-anti-kink summation. Our starting point is the generating functional given in Eq. (\ref{Eq:Z-before-kink}), to be evaluated using summation over kink-anti-kink basis. For example, contribution from $\alpha = 2N + 1$ configuration with all but one $n_i = 1$ is given by a sum over one-kink states. Contribution from a $\alpha(\alpha - 1)/2!$ configurations with all, but two $n_i = 1$ can be represented by a sum over kink-anti-kink states

\beqa
\sum_{k_1 = -N}^N \sum_{k_2 = -N}^{k_1 -1} =\frac{\alpha(\alpha -1)}{2!}. 
\eeqa

Because of the periodic boundary condition, the 1-kink configurations actually correspond to a kink-anti-kink configuration with the anti-kink at the end of the lattice or a set of $\alpha$ anti-kink configurations correspond to anti-kink-kink configurations with the kink at the end of the lattice. Thus, we simply have 
\beqa
\alpha + \frac{\alpha(\alpha -1)}{2!} =\frac{1}{2} (\alpha+1) \alpha 
= \sum_{k_1 = -N}^{N+1} \sum_{k_2 = -N}^{k_1 -1}.
\eeqa

There are $\alpha(\alpha-1)(\alpha-2)/3!$ configurations with all but three $n_i = 1$. These are 2-kink 2-anti-kink configurations with one kink at the end of the lattice. Adding these to 4-kink 4-anti-kink states

\beqa
\sum_{k_1=-N}^N \sum_{k_2=-N}^{k_1-1} 
\sum_{k_3=-N}^{k_2-1} \sum_{k_4=-N}^{k_3-1}
=\frac{\alpha(\alpha-1)(\alpha-2) (\alpha-3)}{4!},
\eeqa
gives

\beqa
\frac{\alpha(\alpha-1)(\alpha-2) }{3!} +
\frac{\alpha(\alpha-1)(\alpha-2) (\alpha-3)}{4!}
=\sum_{k_1=-N}^{N+1} \sum_{k_2=-N}^{k_1-1} 
\sum_{k_3=-N}^{k_2-1} \sum_{k_4=-N}^{k_3-1}.
\eeqa

Thus, the total number of configurations in the kink basis is given by

\beqa
&&
1+\sum_{k_1=-N}^{N+1} \sum_{k_2=-N}^{k_1-1} 
+\sum_{k_1=-N}^{N+1} \sum_{k_2=-N}^{k_1-1} 
\sum_{k_3=-N}^{k_2-1} \sum_{k_4=-N}^{k_3-1}
+\cdots\nn\\
&&= 1+\alpha +\frac{\alpha(\alpha-1)}{2!}
+\frac{\alpha(\alpha-1)(\alpha-2)}{3!}+\frac{\alpha(\alpha-1)(\alpha-2)(\alpha-3)}{4!}+\cdots
=2^{2N+1}.
\eeqa

We start with the 0-kink 0-anti-kink contribution, and observe that in the limit $\sqrt{2}\mu\beta\gg 1$ it gives the following: 

\beqa
\lb  \frac{\mu^3}{\sqrt{\lambda}}  \rb^2  \frac{1}{\sqrt{2}\mu}
\int_{-\frac{\beta}{2}}^{\frac{\beta}{2}} \mbox{d}\tau_1 \int_{-\frac{\beta}{2}}^{\frac{\beta}{2}} \mbox{d}\tau_2\,
e^{-|\tau_1-\tau_2|\sqrt{2}\mu}
=
\lb  \frac{\mu^3}{\sqrt{\lambda}}  \rb^2  \frac{\beta^2}{\sqrt{2}\mu}
\int_{-\frac{1}{2}}^{\frac{1}{2}} \mbox{d}\tau_1 \int_{-\frac{1}{2}}^{\frac{1}{2}} \mbox{d}\tau_2\, e^{-|\tau_1-\tau_2|\sqrt{2}\mu \beta}
\approx\frac{\beta \mu^4}{\lambda}.
\eeqa 

In obtaining the above, we have approximated the discrete site summation (along temporal direction) by continuum integration $a\sum_{site} \to \int d\tau_{i}$ by virtue of the limit $\sqrt{2}\mu\beta\gg 1$ (i.e. very low temperature limit): note this is an approximation we will use throughout the derivation.  Another point is that for the integration over  $e^{-{|\tau_1-\tau_2|\sqrt{2}\mu \beta}}$ we keep only the dominant, leading order contribution that comes from an interval $|\tau_1-\tau_2|< 1/\sqrt{2}\mu \beta$ with its width $\sim 2/\sqrt{2}\mu \beta$.  This point is also to be used repeatedly later. 
The above result is precisely the match of the constant term in the action Eq. (\ref{Eq:Z-before-kink}) albeit with an opposite sign. This motivates us to rewrite the action into the following form:   
\beqa
-\beta \frac{\mu^4}{\lambda} 
+ \lb \frac{\mu^3}{\sqrt{\lambda}} \rb^2 a^2 \sum_{n,m= -N}^N  n_n G(n -m)  n_m  
\to  \lb \frac{\mu^3}{\sqrt{\lambda}} \rb^2 a^2 \sum_{n,m= -N}^N  [n_n
n_m  -1] G(n -m). 
\eeqa

Next, consider the 1-kink 1-anti-kink contribution to the action. Here, without loss of generality, we will assume an initial condition, e.g. $n_{-N+1}=1$ such that when $n=k_1,k_2$, the sign functions are $\mbox{sgn}(k_1-n)=-1$ and $\mbox{sgn}(k_2-n)=1$, respectively. The sum over these configurations can be evaluated as follows: 
\beqa
&&\lb \frac{\mu^3}{\sqrt{\lambda}} \rb^2 a^2 
\sum_{n,m=-N}^N [\mbox{sgn}(k_1-n) \mbox{sgn}(k_2-n) \mbox{sgn}(k_1-m) \mbox{sgn}(k_2-m)-1]G(n -m)\nn\\
&&=\lb \frac{\mu^3}{\sqrt{\lambda}} \rb^2 a^2 
\sum_{n=-N-k_1}^{N-k_1} \sum_{m=-N-k_1}^{N-k_1} 
[\mbox{sgn}(-n) \mbox{sgn}(\tilde{\Delta}-n) \mbox{sgn}(-m) \mbox{sgn}(\tilde{\Delta}-m)-1]G(n - m)\nn\\
&&=\lb \frac{\mu^3}{\sqrt{\lambda}} \rb^2 a^2 
\sum_{n=-N}^{N} \sum_{m=-N}^{N} 
[\mbox{sgn}(-n) \mbox{sgn}(\tilde{\Delta}-n) \mbox{sgn}(-m) \mbox{sgn}(\tilde{\Delta}-m)-1]G(n - m)\nn\\
&&\to \lb  \frac{\mu^3}{\sqrt{\lambda}}  \rb^2  \frac{1}{\sqrt{2}\mu}
\int_{-\frac{\beta}{2}}^{\frac{\beta}{2}} \mbox{d}\tau_1 \int_{-\frac{\beta}{2}}^{\frac{\beta}{2}} \mbox{d}\tau_2\,[\mbox{sgn}(-n) \mbox{sgn}(\tilde{\Delta}-n) \mbox{sgn}(-m) \mbox{sgn}(\tilde{\Delta}-m)-1]e^{-|\tau_1-\tau_2|\sqrt{2}\mu}\nn\\
&&\approx  -2 \frac{ \sqrt{2} \mu^3 }{ \lambda }\lb1-e^{-\sqrt{2}\mu\tilde{\Delta}}\rb \approx  -2 \frac{ \sqrt{2} \mu^3 }{ \lambda }, 
\eeqa 
where $\tilde{\Delta}\gg 1/\sqrt{2}\mu$ is the separation between a kink and an anti-kink. 

Recalling from section \ref{DWKink} that  the size of the kink (in the continuum) is $\Delta\tau\sim1/\sqrt{2}\mu$, meaning that in the ``dilute gas" approximation, where kinks and anti-kinks do not overlap, $\tilde{\Delta}$ must be larger than $2\Delta\tau$. Clearly, the DKM is presented as a dilute gas of kinks/anti-kinks, so when approximating the discrete summation to continuum integration, one has to restrict the kink-anti-kink separation in a similar fashion. Therefore, the above result is obtained, at the lowest order approximation, when $\tilde{\Delta}$ is much larger than $1/\sqrt{2}\mu$, otherwise the kink-anti-kink contribution vanishes. It is not difficult to generalize this procedure to the configurations with  $M$-kinks $M$-anti-kinks, for which the summation gives $-2M \lb\sqrt{2}\mu^3/\lambda \rb$. 

The last step involves doing sums over kinks/anti-kinks configurations. To obtain concrete analytic results, we'd like to resort once again to the continuum limit for replacing the sum (over kinks/anti-kinks' positions) by continuous integrals.
 In doing so, however, one needs a proper measure for the conversion between the discrete sum and the continuum integration. The size of a kink provides the natural ``counting measure'' for such a conversion, which we shall adopt as the measure of integration $a' = 1/\sqrt{2}\mu$, and the replacement takes the form:  
\beqa
&&1+\sum_{k_1=-N}^{N+1} \sum_{k_2=-N}^{k_1 -1}
+\sum_{k_1=-N}^{N+1} \sum_{k_2=-N}^{k_1 -1} 
\sum_{k_3=-N}^{k_2 -1} \sum_{k_4=-N}^{k_3 -1} + \cdots \nn\\
&&\to
1+\frac{1}{a'} \int_{0}^{\beta} \mbox{d}\tau_1 \frac{1}{a'} \int_{0}^{\tau_1} \mbox{d}\tau_2
+\frac{1}{a'} \int_{0}^{\beta} \mbox{d}\tau_1 \frac{1}{a'} \int_{0}^{\tau_1} \mbox{d}\tau_2
\frac{1}{a'} \int_{0}^{\tau_2} \mbox{d}\tau_3 \frac{1}{a'} \int_{0}^{\tau_3} \mbox{d}\tau_4
+ \cdots \, . \nn\\
\eeqa

Defining $F=\exp\lb-\sqrt{2} \mu^3 /\lambda\rb$, the partition function $\mathcal{Z}[j=0]$ of Eq. (\ref{Eq:Z-before-kink}) is given by 

\beqa\label{A.Analityc-Z}
\nonumber
\mathcal{Z}[0] = \lb \sum_{n=0}^\infty  e^{-\beta \sqrt{2} \mu  (n+\frac{1}{2}) } \rb 
\ls  1+\frac{\beta^2}{2! a'^2} F^2  
+\frac{\beta^4}{4! a'^4} F^4 +\frac{\beta^6}{6! a'^6} F^6  +\cdots \rs
= \lb \sum_{n=0}^\infty  e^{-\beta \sqrt{2} \mu  (n+\frac{1}{2}) } \rb 
\cosh(\sqrt{2} \mu \beta F).\\
\eeqa 

Finally, we can compute the correlation function 

\beqa
\langle x(\tau)x(0) \rangle 
&=& \ls \frac{\partial}{a \partial j_{\tau_n}} \frac{\partial}{a \partial  j_{0}}
\ln \mathcal{Z} [j]  \rs_{j=0}\nn\\
&=& \frac{G(\tau)}{2} 
+ \frac{1}{\cosh(\sqrt{2} \mu \beta F)} 
\sum_{\mbox{config}}  \lb \frac{\mu^3}{\sqrt{\lambda}} \rb^2
\int \mbox{d}\tau_1\, \mbox{d}\tau_2\, n(\tau_1) G(\tau_1-\tau) n(\tau_2) G(\tau_2-0) \nn\\
&=& \frac{G(\tau)}{2} 
+ \frac{1}{\cosh(\sqrt{2} \mu \beta F)}  \frac{\mu^2}{\lambda} 
\ls  
1+\frac{1}{a'} \int_0^\beta \mbox{d}\tau_1 \frac{1}{a'} \int_0^{\tau_1} \mbox{d}\tau_2 
~\mbox{sgn}(\tau_1 -\tau)  \mbox{sgn}(\tau_2 -\tau) \mbox{sgn}(\tau_1)
\mbox{sgn}(\tau_2) F^2 
\right. \nn\\
&+& \left. \frac{1}{a'} \int_0^\beta \mbox{d}\tau_1 \frac{1}{a'} \int_0^{\tau_1} \mbox{d}\tau_2 
\frac{1}{a'} \int_0^{\tau_2} \mbox{d}\tau_3 \frac{1}{a'} \int_0^{\tau_3} \mbox{d}\tau_4 
~\mbox{sgn}(\tau_1 -\tau)  \mbox{sgn}(\tau_2 -\tau) \mbox{sgn}(\tau_3 -\tau)\mbox{sgn}(\tau_4 -\tau) \right. \nn\\
&\times& \left.\mbox{sgn}(\tau_1) \mbox{sgn}(\tau_2)  \mbox{sgn}(\tau_3)  \mbox{sgn}(\tau_4) F^4 
+ \cdots   \rs \nn\\
&=& \frac{G(\tau)}{2} +\frac{\mu^2}{\lambda} 
 \frac{\cosh( (\beta-2\tau) \sqrt{2} \mu F)}{\cosh(\sqrt{2} \mu \beta F)},
\eeqa
where

\beqa
\sum_{\mbox{config}} \ls \cdots \rs 
= \lb 1+ \frac{1}{a'} \int_0^\beta \mbox{d}\tau_1 \frac{1}{a'} \int_0^{\tau_1} \mbox{d}\tau_2
[\cdots]  F^2
+\frac{1}{a'} \int_0^\beta \mbox{d}\tau_1 \frac{1}{a'} \int_0^{\tau_1} \mbox{d}\tau_2 
\frac{1}{a'} \int_0^{\tau_2} \mbox{d}\tau_3 \frac{1}{a'} \int_0^{\tau_3} \mbox{d}\tau_4 [\cdots]  F^4
+\cdots  \rb. \nn\\
\eeqa

 \vfil

\end{document}